# Instabilities and shape variation phase transitions in tubular lipid membranes.

*I. Yu. Golushko[1], S. B. Rochal[1], A. Parmeggiani[2] and V. L. Lorman[2]*

[1]*Faculty of Physics, Southern Federal University, 5 Zorge Street, 344090 Rostov-on-Don, Russia*

[2]*Laboratoire Charles Coulomb, UMR 5221 CNRS – Université Montpellier 2, pl. E. Bataillon, 34095 Montpellier, France*

Changes of external parameters in proximity of critical point can increase thermal fluctuations of tubular lipid membrane (TLM) and result in variation of the membrane shape. The phase transitions in the system are shown to be controlled by a single effective parameter, which depends on the pressure difference between inner and outer regions of membrane and the applied stretching force. We determine an interval of the parameter values corresponding to the stability region of the cylindrical shape of TLM and investigate the behavior of the system in the vicinity of critical instabilities, where the cylindrical shape of membrane becomes unstable with respect to thermal fluctuations. The applied boundary conditions strongly influence the behavior of TLM. For example, small negative effective parameter corresponds to chiral shape of TLM only in the case of periodic boundary conditions. We also discuss other three types of phase transitions emerging in the system.

## 1. ВВЕДЕНИЕ

Клеточная мембрана является одним из важнейших компонентов живой клетки. Мембраны играют ключевую роль, как в структурной организации, так и в функционировании всех клеток. Они отделяют клеточное пространство от окружающей среды, участвуют в подавляющем большинстве внутриклеточных процессов, а также соединяют клетки между собой (в тканях). При физиологических условиях большинство липидных мембран проявляют свойства двумерной жидкости, и как следствие способны принимать огромное многообразие форм: от планарных до сферических и цилиндрических.

Рассматриваемые в данной работе трубчатые липидные мембраны (ТЛМ) представляют собой цилиндрические структуры, образованные липидным бислоем и достигающие в длину сотен микрометров при радиусе всего в пару десятков нанометров [1]. ТЛМ участвуют в большинстве внутриклеточных и межклеточных обменных процессов и были обнаружены в целом ряде органелл, таких как митохондрии, комплекс Гольджи и эндоплазматический ретикулум [2,3]. Образование ТЛМ в живых клетках осуществляется моторными белками, которые, перемещаясь вдоль микротрубочек цитоскелета, формируют ТЛМ из клеточных мембран [4].

В связи с развитием экспериментальных методов исследования в последнее время интерес к ТЛМ сильно возрос. Помимо изучения ТЛМ *in vivo*, стало возможным моделирование подобных систем в контролируемых лабораторных условиях [1,4-8]. Достаточно популярным способом получения и исследования ТЛМ является их механическое «вытягивание» из везикул, сформированных либо из биологических мембран, либо из искусственно синтезированных фосфолипидных [1,5-8]. При использовании данного метода объем ТЛМ заметно меняется. До недавнего времени многие физические величины, характеризующие ТЛМ, такие как, например, коэффициент поверхностного натяжения и коэффициент изгибной жесткости мембраны, определялись косвенно при помощи упрощенных моделей, предложенных в 70 годы [9] для качественной интерпретации экспериментальных данных. В моделях, используемых в основном для оценки порядка величины поверхностного натяжения и изгибной жесткости, величиной разности давлений $\Delta P = P_{in} - P_{out}$ между внутренней и внешней областями ТЛМ обычно пренебрегают. Это связано как с малой величиной $\Delta P$, так и с тем, что в таком приближении связь между равновесными значениями коэффициента поверхностного натяжения, изгибной жесткости, радиуса ТЛМ и растягивающей силы $F$ значительно упрощается, вследствие вырождения обобщенного закона Лапласа. Модели другого типа, как например известная модель неустойчивости и динамики ТЛМ [10] вообще не описывают ТЛМ, вытянутую из везикулы. Данные модели пренебрегают силой $F$ и рассматривают ТЛМ при несоответствующем эксперименту условии сохранения ее внутреннего объема. При этом перепад давления $\Delta P$ задается этим условием, а не является внешним параметром. Для полного обзора см. [11].

Недавние эксперименты с использованием современных оптических пинцетов показали, что связь между равновесными параметрами ТЛМ, вытекающая из вырожденного закона Лапласа, не выполняется [12]. Это привело к пониманию того, что модели неустойчивости ТЛМ в приближении $\Delta P=0$ являются неполными и не соответствуют уточненным экспериментальным данным [13,14]. Поэтому построение и исследование адекватных эксперименту теоретических моделей ТЛМ представляется актуальной задачей.

Целью данной работы является исследование модели ТЛМ, находящейся под действием разности давлений между внешней и внутренней областями мембраны и силы, направленной вдоль оси трубки. В большинстве случаев именно в таких условиях находятся как искусственно полученные, так и природные ТЛМ. В данной работе рассматриваются критическое поведение и эволюция формы мембраны в процессе изменения внешних параметров. Исследуются особенности фазовых переходов изменения формы, возможных в ТЛМ и ограничивающих ее область устойчивости. Особое внимание уделено влиянию граничных условий, качественно изменяющих поведение системы.

Статья организована следующим образом. Во втором разделе мы формулируем исследуемую модель и, развивая работу [14], исследуем равновесие цилиндрической формы ТЛМ при разных граничных условиях. Третий раздел посвящен фазовым превращениям в ТЛМ. Заключение содержит основные результаты работы.

## 2. СВОБОДНАЯ ЭНЕРГИЯ МЕМБРАНЫ И РАВНОВЕСИЕ ЕЕ ЦИЛИНДРИЧЕСКОЙ ФОРМЫ

Прежде чем перейти непосредственно к рассмотрению вопросов касающихся равновесной формы мембраны и ее зависимости от внешних условий, следуя работе [14], запишем свободную энергию мембраны в следующем виде:

$$\Phi = \frac{k}{2}\int\left(\frac{1}{R_1}+\frac{1}{R_2}\right)^2 dS + \sigma\int dS - \Delta P\int dV - FL \tag{1}$$

Первый член выражения (1) – свободная энергия Канама-Хелфриха (*Canham-Helfrich free-energy*), где $k$ – изгибная жесткость мембраны, а $R_1$ и $R_2$ главные радиусы кривизны. Второй и третий члены представляют собой вклады поверхностного натяжения и разности давлений в свободную энергию мембраны, здесь $\sigma$ – энергия на единицу площади границы раздела, $\Delta P = P_{in} - P_{out}$ – разность давлений внутри и снаружи ТЛМ. $L$ – длина ТЛМ, $F$ – приложенная вдоль оси ТЛМ растягивающая (если $F>0$) сила, а $dV$ и $dS$ – дифференциалы объема и площади соответственно.

Для того чтобы пояснить физическую модель, описываемую энергией (1), приведем принципиальную схему эксперимента [15] по вытягиванию ТЛМ (см. рис.

1). Использование оптического пинцета для удержания стеклянного шарика [7,8] в такого рода экспериментах позволяет измерять силу $F$, действующую на ТЛМ. Радиус трубки при этом обычно рассчитывается в предположении, что ТЛМ имеет форму правильного цилиндра [1].

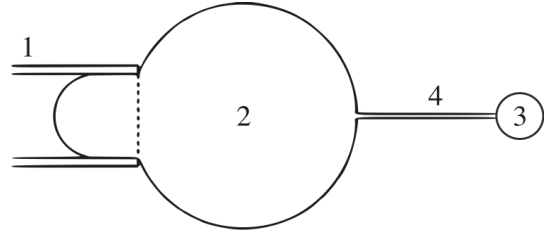

*Рисунок 1. Схема эксперимента по вытягиванию ТЛМ. 1 – микропипетка, 2 – везикула, 3 – стеклянный или полимерный шарик, 4 – ТЛМ [10].*

Чтобы исследовать равновесие формы, будем считать, что мембрана представляет собой цилиндр длиной $L$ и радиусом $r$ с отношением $r/L \ll 1$. Минимизируя свободную энергию такой мембраны по $r$ и $L$, получаем два обобщенных уравнения Лапласа:

$$\sigma = r\Delta P + \frac{k}{2r^2} \; ; \; F = \sigma 2\pi r - \Delta P \pi r^2 + \frac{k\pi}{r} \qquad (2)$$

Таким образом, независимыми являются лишь три из пяти переменных характеризующих равновесную ТЛМ. Например, в экспериментах по вытягиванию трубчатых липидных мембран [7,8], рассмотренных выше, $r$ и $F$ определяются разностью давлений $\Delta P$ при постоянных величинах $\varsigma$ и $k$. Отметим, что условие $\Delta P=0$ приводит к вырождению закона Лапласа и как следствие к более простой зависимости $r$ от $F$.

Для анализа неустойчивостей формы относительно флуктуаций будем рассматривать поверхность мембраны как двумерное многообразие **R**$(x,y,z)$ в 3D пространстве. В цилиндрической системе координат радиальное поле смещений $u_r$ элементов поверхности ТЛМ выражается функцией от угла $\varphi$ и координаты $z$, при этом координаты самого элемента в декартовой системе координат записываются в виде $x=[u_r(\phi,z)+r]\cos\phi$, $y=[u_r(\phi,z)+r]\sin\phi$, $z=z'$. Площадь элемента поверхности определяется через первую квадратичную форму $dS=\sqrt{E_d G_d - F_d^{\,2}}\,drd\phi$, где $E_d = \left(\frac{\partial x}{\partial z}\right)^2 + \left(\frac{\partial y}{\partial z}\right)^2 + 1$, $F_d = \frac{\partial^2 x}{\partial z \partial \phi} + \frac{\partial^2 y}{\partial z \partial \phi}$,

$G_d = \left(\dfrac{\partial x}{\partial \phi}\right)^2 + \left(\dfrac{\partial y}{\partial \phi}\right)^2$, а объем деформированной ТЛМ как $V = \dfrac{1}{2}\int_0^L \int_0^{2\pi}(x^2+y^2)d\phi dz$.

При описании малых флуктуаций можно ограничить разложение свободной энергии в ряд квадратичными членами по полю деформаций $u_r(\phi,z)$ и его производным. Сделав это и выразив $\Delta P$, $S$ через $k$, $F$ и $r$ из (2), мы получаем:

$$\Phi = \int\limits_{z=0}^{L} \int\limits_{\phi=0}^{2\pi} \{kA(\phi,z) + FB(\phi,z) - kC(\phi,z)\} d\phi dz, \qquad (3)$$

где

$$A(\phi,z) = \dfrac{1}{2r^3}\left((\partial_\phi^2 u_r)^2 + 2(\partial_z^2 u_r)(\partial_\phi^2 u_r)r^3 + 3u_r^2 + 4u_r(\partial_\phi^2 u_r) - 2(\partial_z u_r)^2 r^2 + (\partial_z^2 u_r)^2 r^4\right),$$

$$B(\phi,z) = \dfrac{1}{2\pi r^2}\left((\partial_\phi u_r)^2 + (\partial_z u_r)^2 r^2 - 4u_r^2 + r^2\right),$$

$$C(\phi,z) = \dfrac{1}{r^2}\left(\partial_\phi^2 u_r + r^2 \partial_z^2 u_r\right).$$

Для правильного описания флуктуационного поведения мембраны, и как следствие определения пределов ее устойчивости необходимо выбирать граничные условия, адекватные эксперименту, так как выбор граничных условий существенно влияет на поведение модели. Рассмотрим сначала периодические граничные условия. Поле деформаций в таком случае может быть представлено в виде следующей линейной комбинации базисных функций:

$$u_r(\phi,z) = \sum_{n,m=-\infty}^{\infty} A_{n,m} e^{i(n\phi + k_m z)} \qquad (4)$$

где $k_m = 2\pi n/L$ - волновой вектор ТЛМ конечной длины $L$. Подставив разложение (4) в (3), получаем следующее выражение для свободной энергии мембраны:

$$\Phi = \dfrac{\pi L k}{r^3} \sum_{n,m=-\infty}^{\infty} M_{n,m} A_{n,m} A_{n,m}^* \qquad (5)$$

где

$$M_{n,m} = (r^2 k_m^2 + n^2 - 1)^2 - 2(n^2 - 1) + \alpha(r^2 k_m^2 + n^2 - 1), \qquad (6)$$

и $\alpha = \dfrac{Fr}{\pi k}$ (здесь и далее нормализованная сила) является единственным внешним параметром системы. Заметим, что при выборе периодических граничных условий все линейные по полю смещений и его производным члены обращаются в ноль. Цилиндрическая форма очевидно устойчива при тех значениях $\alpha$, для которых коэффициенты $M_{n,m}$ положительны. Несложно получить, что положительным $M_{nm}$ соответствует интервал значений параметра $\alpha$ $(-(2\pi r/L)^2;3)$. На левой границе этого интервала первой обращается в ноль энергия деформации, соответствующая четырем вырожденным модам с $n=\pm 1$ и $m=\pm 1$, а на правой – моде с $n=0$ $m=0$. Также заметим, что в случае с периодическими граничными условиями моды с $n=\pm 1$ и $m=0$ всегда являются безэнергетическими голдстоуновскими (трансляционными). Соответствующие элементы $M_{n,m}$ обращаются в ноль, независимо от величины приложенной силы $\alpha$, что никак не сказывается на устойчивости мембраны.

Исследовать случай ТЛМ с закрепленными концами немного сложнее. Теперь поле смещений должно обращаться в ноль на концах ТЛМ и разложение $u_r(\phi,z)$ приобретает вид:

$$u_r(\phi,z) = \sum_{n=-\infty}^{\infty}\sum_{m=1}^{\infty} e^{in\varphi}\sin(k_m z) \qquad (7)$$

а выражение для свободной энергии мембраны получается в форме:

$$\Phi = \frac{\pi L k}{2r^3}\sum_{n=-\infty}^{\infty}\sum_{m=1}^{\infty} M_{n,m}A_{n,m}A^*_{n,m} + \frac{4k\pi^2}{L}\sum_{m=1,3,5,...}^{\infty} m A_{0,m} \qquad (8)$$

где $k_m = \pi n/L$, а $M_{n,m}$ по-прежнему определяется выражением (6). Условие положительной определенности коэффициентов $M_{n,m}$ несколько иное, и интервал $\alpha$ определяется: слева – обращением в ноль энергии, соответствующей дважды вырожденной моде с $n=\pm 1$ $m=1$, справа – моде $n=0, m=1$, поскольку функция с $m=0$ не входит в разложение (7). Окончательно интервал искомых значений параметра $\alpha$ есть $(-(\pi r/L)^2; 3+\alpha_0)$, где $\alpha_0 = \left(\frac{\pi r}{L}\right)^2\left(\frac{L^2+\pi^2 r^2}{L^2-\pi^2 r^2}\right) << 1$.

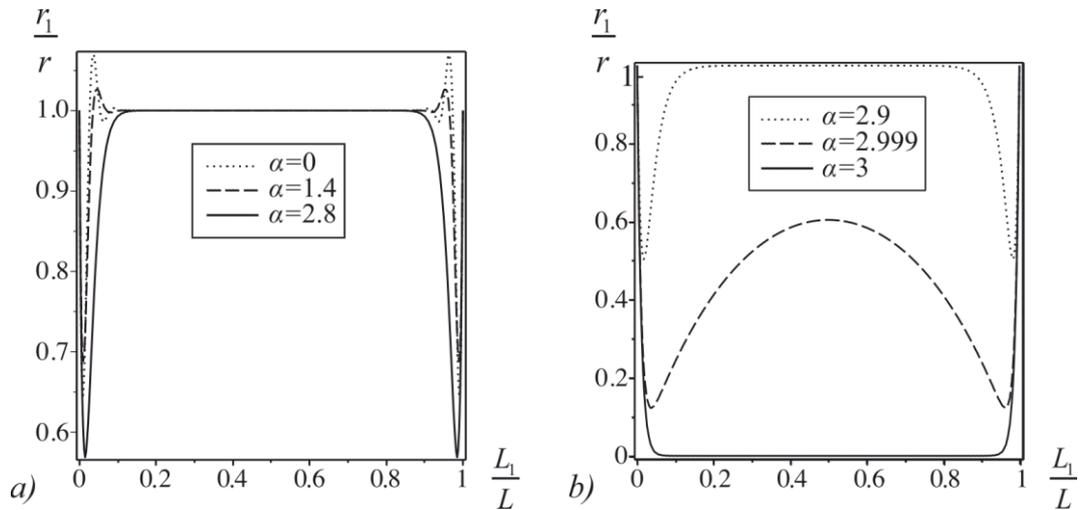

*Рисунок 2. Форма мембраны, описываемой энергией (8) при изменении внешнего параметра α. a) – в интервале [0..2.8] и b) – в интервале [2.9..3]. Фактически при значениях α из первого интервала текущий радиус ТЛМ $r_1$отличается от величины r только вблизи краев трубки.*

Подчеркнем, что наличие линейных членов в (8) свидетельствует о том, что в случае закрепленных концов равновесная форма мембраны не будет строго цилиндрической, поэтому знаки коэффициентов $M_{n,m}$ не могут строго определить область устойчивости мембраны в соответствующей фазе. Однако существует широкий интервал значений *α,* в котором форма мембраны близка к цилиндрической и в котором разложение (8) представляется уместным для анализа малых деформаций мембраны. Минимизация (8) дает следующее выражение для амплитуд гармоник поля равновесных деформаций:

$$A_{0,m} = -\frac{4\pi r^3}{L^2}\frac{m}{M_{0,m}} \qquad (9)$$

где *n=0* и *m=*1,3,5,... При этом мембрана сохраняет ось вращения, а ее форма может зависеть от значения внешнего параметра *α*. Однако, как это видно из рис. 2a в области $0 \leq \alpha \leq 2.8$ форма мембраны близка к цилиндрической. Лишь вблизи границ ТЛМ ее радиус $r_1$ осциллирует около равновесного значения *r*, определяемого условиями равновесия цилиндрической формы (2). При приближении к особой точке *α=3+α₀* минимизация (8) приводит к форме (см. рис. 2b), не имеющей ничего общего с исходной цилиндрической. В окрестности этой точки разложение (8) очевидно является недостаточным и требуется учет старших членов.

## 3. ФАЗОВЫЕ ПРЕВРАЩЕНИЯ ИЗМЕНЕНИЯ ФОРМЫ В ТЛМ

Описание потери устойчивости формы липидной мембраны аналогично формализму теории фазовых переходов. В рамках обычной теории фазовых переходов в кристаллах [16] нормализованная сила *α,* введенная ранее, является единственным внешним параметром системы, а параметры порядка (ПП) (как критические, так и нет) есть средние амплитуды поля смещений $|A_{n,m}|$. Амплитуды среднеквадратичных флуктуаций критических ПП при приближении к точкам фазового перехода стремятся к бесконечности. Из энергий (5) и (8) легко получить величину среднеквадратичных радиальных флуктуаций мембраны $\langle |u_r|^2 \rangle$, имеющую аналогичные особенности вблизи критических значений управляющего параметра $\alpha$. В случае периодических граничных условий она имеет вид

$$\langle |u_r|^2 \rangle = A_0^2 \sum_{n,m=-\infty}^{\infty} \frac{1}{M_{n,m}}, \tag{10}$$

где

$$A_0 = \sqrt{\frac{k_B T r^3}{2k\pi L}}$$

В случае закрепленной на концах ТЛМ выражение для ее среднеквадратичных радиальных флуктуаций $\langle |u_r|^2 \rangle$ отличается от (10) лишь явным видом $k_m = \pi n/L$ в матрице $M_{n,m}$ и тем, что суммирование по $m$ начинается с $m=1$. На рисунке 3 приведены зависимости амплитуд флуктуаций параметра порядка от управляющего параметра $\alpha$, в приближении энергии (8).

Вблизи особых точек области изменения параметра α именно соответствующие критические моды дают основной вклад в энергию мембраны. Пользуясь этой стандартной для теории фазовых переходов идеей, мы в окрестности фазового перехода вместо полной энергии мембраны будем рассматривать её разложение, аналогичное потенциалу Л.Д. Ландау, а в поле смещений оставлять лишь критические моды.

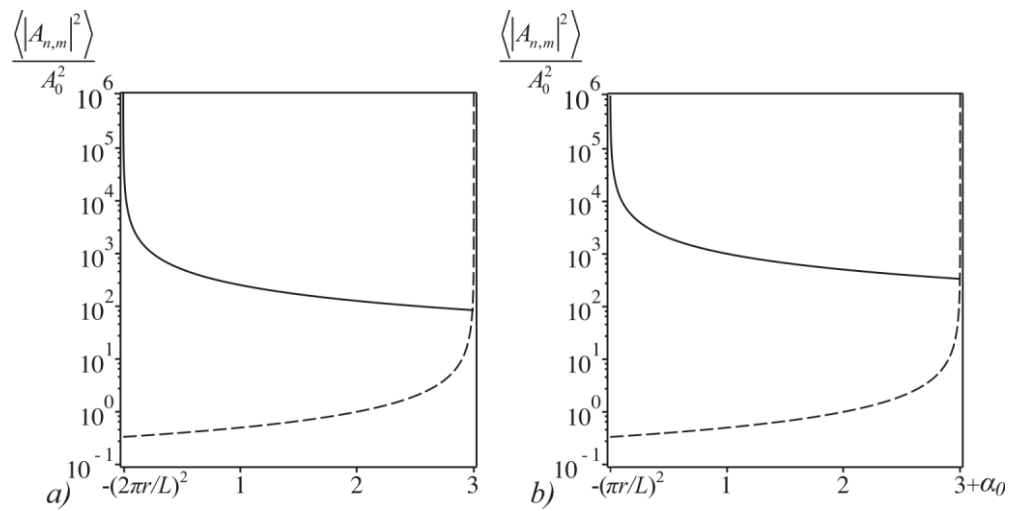

*Рисунок 3. Квадрат приведенной амплитуды среднеквадратичных флуктуаций критических ПП в зависимости от параметра α. а) случай*

*периодических граничных условий: вырожденные моды с n=±1 m=±1 (сплошная линия) и мода с n=0 m=0 (прерывистая линия); b) случай ТЛМ с закрепленными концами: вырожденные моды с n=±1 m=1 (сплошная линия) и мода с n=0 m=1 (прерывистая линия).*

В случае ТЛМ с закрепленными концами, в точке $\alpha=-(\pi r/L)^2$ фазовый переход связан с конденсацией мод с $n=\pm 1$ и $m=1$. Оставляя в поле смещений только эти две моды, удобно привести его к действительному виду (11), в котором вследствие данного преобразования остается только одна амплитуда $A$ и появляется дополнительная фаза $\varphi_0$.

$$u(\phi, z) = A\cos(\phi+\phi_0)\sin\left(\frac{\pi}{L}z\right) \qquad (11)$$

Далее, подставляя поле (11) в исходную энергию (1) с учетом соотношений (2), мы разлагаем подынтегральное выражение в ряд по степеням $A$ и интегрируем полученное разложение. В результате этих преобразований получается приближенное полиноминальное по $A$ выражение для свободной энергии мембраны вблизи фазового перехода. Заметим, что получаемый таким образом потенциал Ландау не зависит от дополнительной фазы $\varphi_0$, соответствующей однородному вращению поля смещений вокруг оси трубки. Ограничимся потенциалом четвертой степени, так как члены старшего порядка качественно не меняют форму зависимости потенциала Ландау от амплитуды $A$:

$$\Phi = A^2 \frac{\pi^3 k}{2rL}\left(\frac{r^2\pi^2}{L^2}+\alpha\right) + A^4 \frac{L\pi k}{32 r^5}\left(3(9+\alpha) - 30\left(\frac{\pi r}{L}\right)^6 + \left(\frac{\pi r}{L}\right)^4(1-9\alpha) + 2\left(\frac{\pi r}{L}\right)^2(1-\alpha)\right) \qquad (12)$$

В выражении (12) отсутствуют как линейные так и кубические по $A$ члены, при этом при переходе через точку $\alpha=-(\pi r/L)^2$ коэффициент при $A^2$ меняет свой знак на противоположный, в то время как коэффициент при $A^4$ остается положительным, как и четные члены старших порядков. Поэтому данное превращение является фазовым переходом второго рода, а в закритической области отрицательных значений нормализованной силы $\alpha$ существует единственная низкосимметричная фаза, соответствующая деформации мембраны вида

$$u(\phi, z) = A\cos(\phi)\sin\left(\frac{\pi}{L}z\right), \qquad (13)$$

представленной на рис. 4. Такое поведение мембраны вполне очевидно и во многом аналогично упругой Эйлеровой неустойчивости [17].

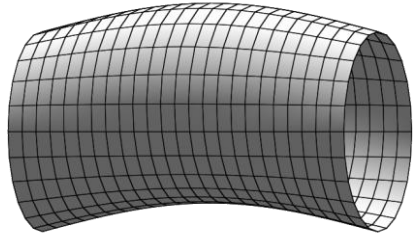

*Рисунок 4. Форма мембраны при небольшой закритической отрицательной величине приведенной силы. Рассматриваются граничные условия с закрепленными концами.*

При рассмотрении ТЛМ с периодическими граничными условиями поведение системы изменяется кардинальным образом. Вблизи точки фазового перехода $\alpha=-(2\pi r/L)^2$ критическое поле деформации мембраны связанное с четырьмя модами $n=\pm 1$ и $m=\pm 1$ может быть представлено в виде действительного выражения

$$u(\phi,z) = A\cos\left(\phi+\phi_0+\frac{2\pi}{L}(z-z_0)\right) + B\cos\left(-\phi+\phi_0+\frac{2\pi}{L}(z-z_0)\right), \qquad (14)$$

где переменные $\varphi_0$ и $z_0$ соответствуют однородному вращению и однородному смещению поля относительно ТЛМ.

Далее, воспользовавшись полем смещений (14) и энергией (1), получаем выражение для потенциала Ландау ТЛМ, которое также можно ограничить членами четвертой степени по компонентам параметра порядка $A$ и $B$:

$$\begin{aligned}\Phi = &(A^2+B^2)\frac{(2\pi)^3 k}{2Lr}\left(\alpha+\left(\frac{2\pi r}{L}\right)^2\right) \\ &+ A^2 B^2 \frac{L\pi k}{2r^5}\left(9+\alpha-10\left(\frac{2\pi r}{L}\right)^6+\left(\frac{2\pi r}{L}\right)^4(43-3\alpha)-10\left(\frac{2\pi r}{L}\right)^2(1-\alpha)\right) \\ &+ (A^2+B^2)^2\frac{L\pi k}{4r^5}\left(9+\alpha-10\left(\frac{2\pi r}{L}\right)^6-3\left(\frac{2\pi r}{L}\right)^4(7+\alpha)+6\left(\frac{2\pi r}{L}\right)^2(1-\alpha)\right)\end{aligned} \qquad (15)$$

Аналогично предыдущему случаю полученный потенциал (15) не зависит от голдстоуновских переменных $\varphi_0$ и $z_0$, а его симметрия запрещает линейные и кубические по компонентам ПП члены. Коэффициент при $A^2+B^2$ меняет свой знак при переходе через точку $\alpha=-(2\pi r/L)^2$, в то время как коэффициенты при $A^2B^2$ и $(A^2+B^2)^2$ остаются существенно положительными. Таким образом, как и в случае закрепленной на концах ТЛМ, рассматриваемый переход является фазовым переходом второго рода. Однако у ТЛМ с периодическими граничными условиями в закритической области существует два типа решений уравнения состояния: $A\neq 0$, $B=0$ (хиральное) и $A=B$ (ахиральное, более симметричное), энергия которого очевидно повышается отсутствующим в предыдущем случае членом при $A^2B^2$. Кроме того решение $A=B$ является еще и неустойчивым. Таким образом, в ближней закритической области при $\alpha<-(2\pi r/L)^2$ существует единственная фаза $A\neq 0$, $B=0$. В этом состоянии ТЛМ приобретает хиральную спиралевидную форму. Так как остальные некритические ПП малы, то хиральная форма (см. рис. 5) деформированной трубки приближенно задается уравнением:

$$u(\phi,z) = A\cos\left(\phi + \frac{2\pi}{L}z\right) \quad (16)$$

Рассмотрим теперь противоположную границу устойчивости цилиндрической ТЛМ, т.е. другой предельный случай, соответствующий положительной нормализованной силе с величиной, принадлежащей закритической области. Как уже было упомянуто выше, в случае периодических граничных условий правая граница области устойчивости мембраны определяется модой $n=0$ $m=0$, соответствующей изменению радиуса ТЛМ. Ввиду простоты выражения для деформации мембраны вблизи этой критической точки ($u(\phi,z) = A$), отпадает необходимость разложения энергии в ряд, и она приобретает вид:

$$\Phi = A^2 \frac{L\pi k}{2(A+r)r^3}\left(A(2-\alpha)+r(3-\alpha)\right) \quad (17)$$

Рассматривать данное выражение имеет смысл лишь при условии $A>-r$, так как величина деформации сжатия по абсолютной величине не может превышать радиус ТЛМ.

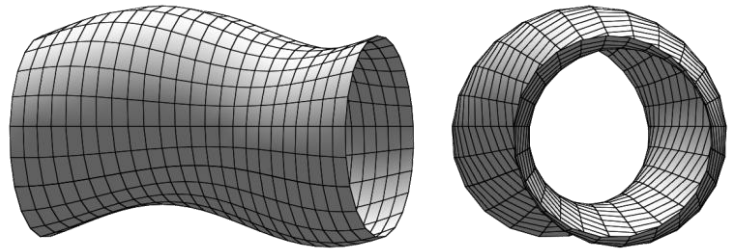

*Рисунок 5. Хиральная форма мембраны при небольшой закритической отрицательной величине приведенной силы. Рассматриваются периодические граничные условия.*

Анализ (17) показывает, что при величине *α*, лежащей в интервале *(0;2)*, система устойчива и имеет глобальный минимум *A*=0 (см. рис. 6). При дальнейшем увеличении внешней силы (2<*α*<3) минимум *A*=0 становится локальным, а соответствующая ему фаза становится метастабильной. Глобальный минимум в этой области параметра *α* отсутствует. При *α*=3 фаза *A*=0 теряет устойчивость, после чего ТЛМ либо вообще разрывается (потеряв устойчивость), либо, претерпев фазовый переход второго рода, попадает в появляющийся при *α*>3 локальный минимум A<0, соответствующий ТЛМ меньшего радиуса (см. рис. 6). В последнем случае формально диаметр ТЛМ обращается в 0 лишь при бесконечно большой нормализованной силе. Однако формирование ТЛМ *in vivo* или *in vitro* скорее сопровождается постепенным возрастанием параметра *α* нежели, его скачкообразным изменением. Будь это формирование связано с ростом растягивающей силы *F,* увеличением разности давлений *∆P* [1], или же изменением поверхностного натяжения [18]. Поэтому система, по всей видимости, всегда проходит через ряд промежуточных состояний, в которых термических флуктуаций свободной энергии может быть достаточно для того, чтобы преодолев потенциальный барьер, выйти из локального минимума энергии. Такой процесс при *α*≈3 в свою очередь приведет к разрыву ТЛМ. Поэтому наблюдать реальную метастабильную систему в закритической области представляется весьма проблематичным.

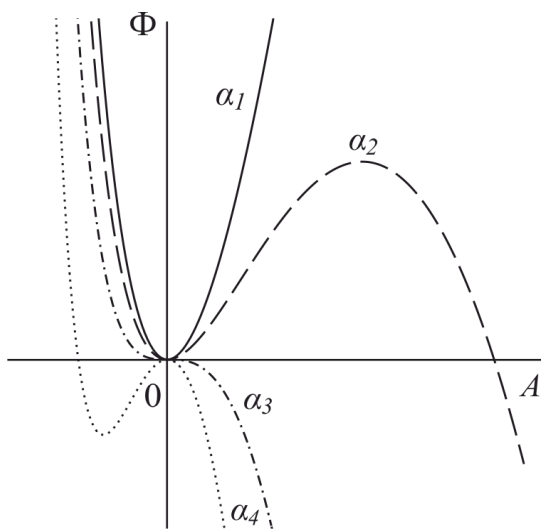

*Рисунок 6. Зависимость потенциала Ландау системы от амплитуды A для различных значений нормализованной силы α: 0<α₁<2, 2<α₂<3, α₃=3, α₄>3. Амплитуда A в данном случае равна изменению радиуса трубки.*

В отличие от поведения трубки с периодическими граничными условиями, описываемого свободной энергией (5), поведение ТЛМ с зажатыми концами вблизи правой границы интервала устойчивости невозможно свести к исследованию энергии (8). В этой области роль других некритических ПП с $n=0$ и $m>1$ намного существеннее, чем в области $α<3$. Описание таких сложных неустойчивостей сильно отличается от представленного выше, и требует уровня математического анализа проблемы, выходящего за рамки настоящей работы. Отметим только, что при помощи подобного детального анализа, возможно, удастся объяснить появления так называемой жемчужной (pearling) нестабильности [19,20], при которой на поверхности ТЛМ формируется серия вздутий.

## 4. ЗАКЛЮЧЕНИЕ

Построение и развитие адекватных эксперименту физических моделей липидных мембран является актуальной задачей. В настоящей работе, развивающей модель ТЛМ с переменным внутренним объемом [14], впервые исследованы неустойчивости и фазовые переходы изменения цилиндрической формы ТЛМ, «вытянутой» из везикулы. Для расчета формы мембраны в слабозакритических областях, мы предположили, что основной вклад в ее изменение дают именно критические флуктуации, что позволило учесть необходимые старшие члены разложения свободной энергии ТЛМ. Предложенный подход, основанный на теории фазовых переходов Л.Д. Ландау в кристаллах и концепции критических параметров порядка, выгодно отличает настоящую работу от предшествующих работ, также посвященных исследованию стабильности ТЛМ [14,15,21].

В работе также показано, что граничные условия существенно сказываются на критическом и слабозакритическом поведении ТЛМ. В частности, ТЛМ с закрепленными концами имеет большую область устойчивости, чем ТЛМ, рассматриваемая в приближении периодических граничных условий. Проанализировано четыре варианта возможной неустойчивости высокосимметричной фазы, соответствующих комбинациям двух типов граничных условий (периодических и закрепленных концов) и двум типам критических флуктуаций, возможных в системе. На одной из границ устойчивости в закрепленной на концах мембране возникает упругая неустойчивость Эйлерова типа [17], и мембрана деформируется подобно классическому стержню, к которому приложена критическая сжимающая сила. При близкой величине $α$, но при периодических граничных условиях, рассматриваемая цилиндрическая мембрана, претерпевая другой фазовый переход второго рода, приобретает хиральную спиралевидную форму. Исследование системы при втором особом значении $α≈3$ показывает, что критические флуктуации второго типа, индуцируемые растяжением мембраны, возможны лишь при периодических граничных условиях и ведут к разрыву мембраны.

## Список литературы